\documentclass[a4paper,11pt]{article}
\usepackage[utf8]{inputenc}
\usepackage[english]{babel}
\usepackage{amsmath,amssymb,amsthm,indentfirst}
\usepackage[big]{layaureo}

\theoremstyle{definition}

\newtheorem{example}{Example}
\newtheorem{definition}{Definition}

\title{An Introduction to Imperfect Competition\\ via Bilateral Oligopoly}
\author{Alex Dickson\thanks{Department of Economics, University of Strathclyde, Glasgow, G1 1XQ, UK.} \and Simone Tonin\thanks{Durham Business School, Durham University, Durham, DH1 3LB, UK.}}
\date{\today}

\begin{document}
\maketitle
\begin{quotation}``\textit{It seems impossible to go on with analysing markets under the assumption of perfect competition. Direct observation of economic activity reveals that markets are the fields of ``giants", operating simultaneously with a fringe of small competitors. Even partial analysis has taken this picture of the market when proposing oligopoly solutions to describe the outcomes of imperfectly competitive markets. Behind the demand function there is a myriad of ``small" price-taking agents, while the supply side is occupied by few agents appearing as giants, contrasting with the dwarfs on the demand side.}"\par \hfill Gabszewicz (2013)\end{quotation}

\section{Introduction}
The study of market power in imperfectly competitive markets has commanded much attention from economists. The workhorse model of industrial organization economists---that of Cournot competition---takes a partial equilibrium perspective and makes the assumption that only firms have market power. Bilateral oligopoly is a natural generalization of Cournot competition to consider markets in which both sellers and buyers can have market power and so behave strategically in manipulating prices to be more favourable to them. The purpose of this exposition is to introduce the reader to the study of bilateral oligopoly by leading them through worked examples that illustrate some of the main ideas to emerge from the literature.\par

Bilateral oligopoly  was first introduced by Gabszewicz and Michel (1997). It consists of an economy in which there are two commodities and agents have ``corner endowments'', i.e. they are endowed with only one of these two commodities. In a noncooperative game, agents can then choose the amount of their initial endowment to put up in exchange for the other commodity. It is often convenient to think of the first of the two commodities as a consumption good and the second as commodity money (which can be viewed as a numeraire), in which case the agents endowed with the consumption good are called sellers, and those endowed with the commodity money are called buyers. Despite the simplicity of the model, the structure is rich enough to study many interesting phenomena that may arise in imperfectly competitive economies. A fruitful line of research focuses on testing the robustness of the partial equilibrium analysis of Cournot oligopolies to general equilibrium models without production where all agents, sellers and buyers, have market power. For instance, Bloch and Ghosal (1997) and Bloch and Ferrer (2001a) study agents' incentives to create submarkets as the Cournot game suggests that firms could gain market power by forming smaller submarkets. Dickson and Hartley (2008) and Amir and Bloch (2009) focus on the existence of Nash equilibrium and the comparative static properties of the equilibrium. Dickson (2013a) pays particular attention to the effect of entry of additional sellers in bilateral oligopoly, comparing the results to the conventional wisdom from the Cournot model.\par

The bilateral oligopoly model belongs to the line of research on strategic market games initiated by the seminal papers of Shubik (1973), Shapley (1976), and Shapley and Shubik (1977). There are many types of strategic market games (see Giraud (2003) and Levando (2012) for a survey). Here we mention only the ``trading post model" and the ``window model" which can be seen as different institutional mechanisms through which prices are determined. In the first model, trade is decentralised through a system of trading posts where commodities are exchanged. Dubey and Shubik (1978) studied the trading post model where only commodity money is used to buy other commodities while Amir, Sahi, Shubik, and Yao (1990) considered the case in which any commodity can be used to buy other commodities.\footnote{A strategic market game with trading post and fiat money was considered by Peck, Shell, and Spear (1992).} Differently, Sahi and Yao (1989) studied the window model where trade is centralised by a clearing house in which there is a ``window" for each commodity. This model was first proposed informally by Lloyd S. Shapley, and we call it the ``Shapley window model". It is important to stress that when there are only two commodities, as in bilateral oligopoly, the three models coincide.\par

The aim of this paper is threefold. First, we provide a unified framework, by means of non-trivial examples,\footnote{Dickson (2013b) shows that examples with Cobb-Douglas utility functions, very common in the literature, have very peculiar features which do not hold in general.} to compare the results obtained in simultaneous-move and sequential-move versions of bilateral oligopoly with the Cournot model and the Cournot-Walras approach. We also underline the advantages and disadvantages of each approach and we survey the main contributions on imperfect competition and strategic market games that are linked with our analysis. Secondly, we show how the bilateral oligopoly model can be used to study different kinds of oligopoly: symmetric oligopoly, where all agents act strategically; asymmetric oligopoly where only some agents act strategically; simultaneous oligopoly, where sellers and buyers make their choices without knowledge of others' decisions; and sequential oligopoly, where some agents move first. Thirdly, we focus on how the bilateral oligopoly model can clarify how either strategic or competitive behaviours may emerge endogenously depending on the fundamentals of the economy. In other words, we show how the bilateral oligopoly model can provide ``a foundation for perfect competition" and ``a foundation for Cournot oligopoly" in two-commodity exchange economies. These are research programs whose aim is to define models where perfect competition and oligopoly are not assumed \textit{a priori} but they are an equilibrium outcome.\footnote{See Gale (2000) for a discussion on why the study of foundations is important.}\par

We start our analysis by recasting the classical Cournot game as an exchange game where firms are replaced by sellers characterised by initial endowments and utility functions. Next, we consider the Cournot-Walras equilibrium concept in exchange economies, introduced by Codognato and Gabszewicz (1991) and Gabszewicz and Michel (1997). This latter model describes, in a general equilibrium framework, the same kind of imperfect competition as the Cournot game. The exchange versions of these cornerstone models to study imperfect competition allow us to compare them with the bilateral oligopoly model in a clear way. In fact all examples in the paper are based on the same exchange economy where sellers have quasi--linear utility functions and buyers have quadratic utility functions that generate a linear demand. By comparing the different approaches we find that there are three main differences between the Cournot-Walras approach and the bilateral oligopoly model. First, the latter model, with a finite number of agents, describes a symmetric oligopoly, since all agents are allowed to act strategically, whereas the Cournot-Walras approach describes an asymmetric oligopoly where only the sellers are allowed to act strategically---they can manipulate prices by changing their actions--- while the buyers are assumed to treat prices as given and beyond their control. Second, the Cournot-Walras approach has an intrinsic two-stage nature which cannot be reconciled with the Cournot-Nash equilibrium of simultaneous-move bilateral oligopoly where all agents act together. In order to capture the two-stage structure considered in Cournot-Walras, we define a sequential bilateral oligopoly, where sellers move in the first stage and buyers move in the second stage, and we adopt a subgame-perfect Nash equilibrium as the equilibrium concept. Third, in bilateral oligopoly all agents are treated symmetrically and no \textit{a priori} assumption is made on agents' behaviour while in Cournot-Walras sellers are assumed to act strategically and buyers are assumed to behaved competitively. Okuno, Postwaite and Roberts (1981) stressed the fact that in those kind of models no explanation is given as to why a particular agent should behave strategically rather than competitively.\par

A natural question to ask is whether it is possible to use the bilateral oligopoly model to study an asymmetric oligopoly where no assumptions on agents' behaviours are made. This is the so-called problem of the ``foundations of oligopoly". We answer positively to that question by showing how a partial replica of the exchange economy --we replicate only buyers-- can be used to obtain different agents' behaviours endogenously. In fact, even if all agents have \textit{a priori} the same strategic position, at the limit of the partial replica sellers turn out to have influence on prices whereas buyers turn out to be competitive. This is due to the fact that buyers lose their ability to influence prices when their number increases. We also show an example where, when only buyers are replicated, sellers' and buyers' commodity bundles at the limit of the subgame-perfect Nash equilibrium of a sequential-move game in which the sellers move first are the same commodity bundles at the Cournot-Walras equilibrium. Therefore, we conclude that the bilateral oligopoly model is useful to study asymmetric oligopoly by using the partial replica. Furthermore, the partial replica in the sequential bilateral oligopoly can provide a foundation of the Cournot-Walras approach.\par

In the last part of our paper, we study the relationship among the Walras (competitive) equilibrium and the other equilibrium concepts examined. To this end, we show an example where, when all agents in the economy are replicated, sellers' and buyers' commodity bundles at the limit of the Cournot-Nash equilibrium correspond to the commodity bundles at the Walras equilibrium. This is not surprising because, as in the partial replica, when an agent is replicated they lose market power and in the limit he behaves competitively. Since all agents are replicated, in the limit everyone behaves competitively and then the equilibrium outcome must be Walrasian. Therefore, the bilateral oligopoly model can be used to study the foundations of perfect competition as competitive behaviours can be obtained endogenously in equilibrium.\par

We conclude our analysis by making some welfare considerations on the different types of competition analysed. By the First Welfare Theorem, the Walras equilibrium is Pareto efficient while the allocations obtained with the other equilibrium concepts are Pareto inefficient. This can easily be seen because there exists an allocation that Pareto dominates the one obtained at any equilibrium different from the Walras equilibrium. In other words, when markets are imperfectly competitive the equilibrium outcome is Pareto inefficient.\par


The rest of the paper has the following structure. In Section 2 we introduce the Cournot exchange game. In Section 3 we describe the Cournot-Walras approach. In Section 4 after having defined the simultaneous bilateral oligopoly model and the Cournot-Nash equilibrium, we give an example with a finite number of agents and an example where we partially replicate the exchange economy concluding there may be differences between a simultaneous bilateral oligopoly and the Cournot-Walras approach even when the number of buyers is large. In Section 5 we introduce the sequential bilateral oligopoly model, we partially replicate the exchange economy, and we compare the limit of the subgame-perfect Nash equilibrium with the Cournot-Walras equilibrium. In Section 6 we compare our previous results with the Walras equilibrium and we make some welfare considerations. In section 7 we draw some conclusions and we outline some open problems.

\section{Cournot exchange game (partial analysis)}
 \begin{quotation}
``Let us now imagine two proprietors and two springs of which the qualities are identical, and which, on account of their similar positions, supply the same market in competition."\par \hfill \emph{A. A. Cournot (1838)}\end{quotation}\par

In this section, by taking a slightly different approach from the one proposed by Cournot (1838), we consider the two proprietors as sellers of water and not as producers. This means that they are characterised by utility functions instead of cost functions. The interpretation, however, is very similar: supplying some of the good to the market reduces the seller's holdings of it that provides some disutility, i.e. a cost. The consumption good (water in Cournot's example) is denoted by $x$ and it is exchanged for money which is denoted by $y$.\footnote{Commodity $y$ should be called commodity money because it enters in the utility function. However, for simplicity, we simply refer to it as money.} A commodity bundle $(x,y)$ is a point in $\mathbb{R}_+^2$ which is the set of all feasible commodity bundles. As in Cournot (1838), we assume that buyers are represented by the demand for the consumption good which is a downward sloping function of the price, i.e., $D=f(p_x)$. Each seller is characterised by a utility function, $u_i:\mathbb{R}_+^2\rightarrow \mathbb{R}$, which represents their preferences, and by an initial endowment, $(x_i^0,y_i^0)\in \mathbb{R}_+^2$, such that $x_i^0>0$ and $y_i^0=0$, i.e., sellers hold only the consumption good. Sellers face the demand and we assume that they choose a supply of the consumption good in order to obtain a commodity bundle which maximises their utility. We suppose that there are $m$ sellers.\par

We now introduce the Cournot exchange game $\Gamma$. The strategy set of seller $i$ is
\begin{equation}\label{CE:sellerstrategy}
\mathcal{Q}_i=\{q_i\in \mathbb{R}:0\leq q_i\leq x_i^0\},\end{equation}
with $q_i$ the offer of the consumption good that seller $i$ puts up in exchange for money. Let $\mathcal{Q}=\prod _{i=1}^m\mathcal{Q}_i$ and $\mathcal{Q}_{-h}=\prod _{i\neq h}\mathcal{Q}_i$. Let $q=(q_1,\dots,q_m)$ and $ q_{-i}=(q_1,\dots,q_{i-1},q_{i+1},\dots,q_m)$ be elements of $\mathcal{Q}$ and $\mathcal{Q}_{-i}$ respectively. We denote by $Q=\sum_{i=1}^mq_i$ the total amount of the consumption good sold by sellers. We assume that the market for the consumption good clears, which means that the total supply is equal to the demand. Therefore, for each $q\in \mathcal{Q}$, the price of the consumption good is $p_x(q)=f^{-1}(Q)$ which implies $D=Q$ and $p_y$ is normalized to 1. For each $q\in\mathcal{Q}$, the commodity bundle $(x_i(q),y_i(q))$ of a seller $i$ is given by
\begin{equation}\label{CE:allocation}\begin{aligned}
x_i(q)&=x_i^0-q_i,\\
y_i(q)&=p_x(q)\cdot q_i,\end{aligned}\end{equation}
for $i=1,\dots,m$. The payoff function of a seller $i$ is
\begin{equation}\label{CE:sellerpayoff}
\pi_i(q)=u_i(x_i(q),y_i(q)),\end{equation}
for $i=1,\dots,m$. A Cournot exchange game is then a set $\Gamma=\{(\mathcal{Q}_i,\pi_i(\cdot))_{i=1}^m\}$. We now introduce the definition of a Cournot equilibrium.
\begin{definition}The strategy profile $\hat{q}$ is a Cournot equilibrium for the game $\Gamma$ if for each seller $i=1,\ldots,m$ we have $\pi_i(\hat q_i,\hat q_{-i})\geq\pi_i(q_i,\hat q_{-i})$, for each $ q_i\in \mathcal Q_i$.\end{definition}

We now consider an example to illustrate the Cournot equilibrium concept.
\begin{example}Consider a market with two sellers such that $u_i(x,y)=\log(1+x)+y$ and $(x_i^0,y_i^0)=(3,0)$, for $i=1,2$. Buyers are represented by the demand function $D=6-2p_x$. Since we assume that the market for the consumption good clears, $D=q_1+q_2$, we obtain $p_x(q)=3-\frac{1}{2}(q_1+q_2)$. To find the Cournot equilibrium we need to find the strategies $\hat{q}_1$ and $\hat{q}_2$ which maximise the sellers' payoffs. Consider, without loss of generality, the maximization problem for seller 1
\begin{equation}\label{CE:max}\begin{aligned}
&\underset{q_1}{\text{max}}& & \log(1+(3-q_1))+\Bigl(3-\frac{1}{2}(q_1+q_2)\Bigr)q_1,\\
& \text{subject to}& & 0\leq q_1\leq 3.
\end{aligned}\end{equation}
This payoff function is strictly concave in $q_1$, and an easy way to find the maximum is to solve the problem as an unconstrained maximisation problem then check the constraints are satisfied. The first-order condition is
$$\frac{\partial \pi_1}{\partial q_1}:-\frac{1}{4-q_1}+3-q_1-\frac{q_2}{2}=0.$$
Since sellers are identical, we consider the symmetric Cournot equilibrium where $\hat{q}_1=\hat{q}_2$. We then find the following Cournot equilibrium
$$(\hat{q}_1,\hat{q}_2)=\biggl(\frac{9-\sqrt{15}}{3},\frac{9-\sqrt{15}}{3}\biggl).$$
At the Cournot equilibrium the price is $p_x(\hat{q})=\frac{\sqrt{15}}{3}$, demand for the consumption good is $D=\frac{18-2\sqrt{15}}{3}$ and the sellers' commodity bundles are
$$(x_i(\hat{q}),y_i(\hat{q}))=\biggl(\frac{\sqrt{15}}{3},\frac{3\sqrt{15}-5}{3}\biggr),\mbox{ for } i=1,2.$$\qed\end{example}

While we consider a slight variation on the original Cournot model as we treat firms as sellers with utility functions, the underlying principles are exactly the same. Shubik (1973) raised the following critique to this type of oligopoly model:
\begin{quotation}``The Cournot duopoly model is an open market model involving money. After trade has taken place neither the amount of goods nor the amount of money in the system is conserved. Goods flow out into the market and money flows in from the market."\end{quotation}

Shubik's paper advocates a closed market model, where commodities flow within the system and after trade the total amount of each commodity does not change. In our example due to our treatment of the sellers the amount of the consumption good is preserved, but it is immediate to see that the final amount of money is coming from outside the model. This is a typical feature of partial equilibrium models where each market is considered in isolation from the rest of the economy. Therefore, M. Shubik is suggesting to study oligopoly within a general equilibrium model. Furthermore, another critical feature of the Cournot exchange game is that agents are modelled in different ways: sellers are represented by utility functions and initial endowments while buyers are represented by a demand function.\par

In the next section we address these two issues by introducing a two-commodity exchange economy with corner initial endowments and by using the Cournot-Walras equilibrium as solution concept.

\section{Cournot-Walras approach}
We start by defining the buyers who have been represented by a demand function in the previous model. A buyer is characterised by a utility function $u_i:\mathbb{R}_+^2\rightarrow \mathbb{R}$ which represents their preferences and by an initial endowment, $(x_i^0,y_i^0)\in\mathbb{R}_+^2$, such that $x_i^0=0$ and $y_i^0>0$, i.e., buyers hold only money. We suppose that there are $n$ buyers in the exchange economy indexed $i=m+1,\ldots,m+n$. The difference between sellers and buyers in this model lies in the initial endowments: sellers hold only the consumption good while buyers hold only money. We can now formally define an exchange economy $\mathcal{E}=\{(u_i(\cdot),(x_i^0,y_i^0))_{i=1}^{m+n}\}$ which is the set containing all the pairs $(u_i(\cdot),(x_i^0,y_i^0))$ describing sellers and buyers. In two-commodity exchange economies the price vector is simply $p=(p_x,p_y)$.\par

We now describe the agents' behaviours in the Cournot-Walras approach. In the Cournot game sellers were permitted to act strategically while the buyers behind the demand function are implicitly assumed to act as price takers. The Cournot-Walras approach preserves this asymmetry but derives the Walrasian (competitive) demands of the buyers from their characteristics. The Walrasian demands of a buyer $i$ are the functions $x_i(p)$ and $y_i(p)$ that associate to each positive price vector an amount of commodity $x$ and $y$ that maximises the utility function $u_i(\cdot)$ in $i$'s budget set $\{(x,y)\in \mathbb{R}_+^2: p_xx+p_yy\leq p_yy_i^0\}$.\footnote{More generally, Walrasian demands can be correspondences which associate to each price the set of commodity bundles which maximise the utility function in the budget set. For the sake of simplicity, we just consider the case in which Walrasian demands are functions.} In other words, for any positive price vector $p$, the amounts $x_i(p)$ and $y_i(p)$ solve the following maximization problem\footnote{Note that this maximisation problem has a solution only for continuous utility functions and positive prices (see Varian (1992) p. 98).}
\begin{equation*}\begin{aligned}
&\underset{(x,y)\in\mathbb{R}_+^2}{\text{max}}& & u_i(x,y),\\
& \text{subject to}& & p_xx+p_yy\leq p_yy_i^0,\end{aligned}\end{equation*}
for any buyer $i$. We assume again that the market of each commodity clears. Therefore, for each $q\in \mathcal{Q}$, the price vector $p$ at a Cournot-Walras equilibrium must solve
\begin{equation}\label{CWE:price}
\sum_{i=m+1}^{m+n}x_i(p)=\sum_{i=1}^mq_i,\end{equation}
i.e., buyers' total demand of the consumption good must be equal to sellers' total supply of the consumption good. We denote by $p(q)$ the price vector $(p_x(q),1)$ that solves equation (\ref{CWE:price}).\footnote{To avoid cumbersome notation, but without confusion, in the paper we denote the same elements of different models with the same symbol.} If equation (\ref{CWE:price}) holds then the market for money also clears. By having described the price formation rule, which is purely Walrasian, we can now define the rules to calculate the commodity bundle of each agent. The buyers' commodity bundles are given by the Walrasian demands calculated at the price vector $p(q)$. Differently, since sellers' strategies are the same of the Cournot exchange game, their commodity bundles are given according to the equations in (\ref{CE:allocation}) at the price vector $p(q)$. Consequently sellers' payoff functions are defined as in (\ref{CE:sellerpayoff}). We finally introduce the notion of an allocation $(x_i,y_i)_{i=1}^{n+m}$ which is a specification of a commodity bundle for each seller and buyer.\par
We now have all the elements to define the Cournot-Walras equilibrium for the exchange economy $\mathcal{E}$.
\begin{definition}A Cournot-Walras equilibrium for the exchange economy $\mathcal{E}$ is a vector $\tilde{q}$ and an allocation $(\tilde{x}_i,\tilde{y}_i)_{i=1}^{n+m}$ such that
\begin{itemize}
\item[-] for each seller $i=1,\ldots,m$ we have $u_i(x_i(\tilde q_i,\tilde q_{-i}),y_i(\tilde q_i,\tilde q_{-i}))\geq u_i(x_i(q_i,\tilde q_{-i}),y_i(q_i,\tilde q_{-i}))$, for each $q_i\in \mathcal Q_i$;
\item[-] $(\tilde{x}_i,\tilde{y}_i)=(x_i(\tilde{q}),y_i(\tilde{q}))$ for the sellers $i=1,\dots,m$ and $(\tilde{x}_i,\tilde{y}_i)=(x_i(p(\tilde{q})),y_i(p(\tilde{q})))$ for the buyers $i=m+1,\dots,n$.\end{itemize}\end{definition}
We now consider an example to illustrate the Cournot-Walras equilibrium concept.
\begin{example}Consider an exchange economy with 2 sellers and 2 buyers such that $u_i(x,y)=\ln (1+x)+y$ and $(x_i^0,y_i^0)=(3,0)$, for $i=1,2$, and $u_i(x,y)=3x-\frac{1}{2}x^2+y$ and $(x_i^0,y_i^0)=(0,5)$, for $i=3,4$. To calculate a Cournot-Walras equilibrium, the first step is to finding buyers' Walrasian demands. We then solve the constrained maximization problem
\begin{equation*}\begin{aligned}
&\underset{(x,y)\in\mathbb{R}_+^2}{\text{max}}& & 3x-\frac{1}{2}x^2+y,\\
& \text{subject to}& & p_xx+p_yy\leq p_y 5,
\end{aligned}\end{equation*}
for both buyers. By using the Lagrange multiplier method, we can obtain the Walrasian demands
\begin{align*}
&x_i(p)=3-\frac{p_x}{p_y},\\
&y_i(p)=\frac{5p_y^2+p_x^2-3p_xp_y}{p_y^2},
\end{align*}
for $i=3,4$. Study of the second-order conditions reveals the solution corresponds to a maximum.\footnote{It is worth noting that, since the utility functions are quasi-linear, there can be corner solutions. However, for the sake of simplicity and since this does not affect our analysis, we just consider interior solutions.} Since we normalise $p_y=1$, the price of the consumption good which solves equation (\ref{CWE:price}) is
$$p_x(q)=3-\frac{1}{2}(q_1+q_2).$$
By construction, this is the same inverse demand function as in Example 1, but derived from buyers' characteristics rather than assumed. Therefore, it is immediate to see that sellers' maximisation problem is equivalent to the maximisation problem (\ref{CE:max}) in Example 1, so $(\tilde{q}_1,\tilde{q}_2)=(\frac{9-\sqrt{15}}{3},\frac{9-\sqrt{15}}{3})$ and $p(\tilde{q})=(\frac{\sqrt{15}}{3},1)$. Finally, the allocation at the Cournot-Walras equilibrium is
\begin{align*}
(x_i(\tilde{q}),y_i(\tilde{q}))&=\biggl(\frac{\sqrt{15}}{3},\frac{3\sqrt{15}-5}{3}\biggr),\mbox{ for } i=1,2,\\
(x_i(p(\tilde{q})),y_i(p(\tilde{q})))&=\biggl(\frac{9-\sqrt{15}}{3},\frac{20-3\sqrt{15}}{3}\biggr), \mbox{ for } i=3,4.\end{align*}\qed\end{example}
It is immediate to see that sellers' commodity bundles and the total demand of the consumption good from buyers correspond to the ones at the Cournot equilibrium. Therefore, the Cournot-Walras equilibrium captures, in a general equilibrium framework, the same kind of competition as the Cournot exchange game but where the total quantities of the two commodities in the system are preserved. In other words, the Cournot-Walras approach allows us to study asymmetric oligopoly in a general equilibrium setting, with sellers acting strategically and buyers assumed to treat prices as fixed. Therefore, oligopoly models based on exchange economies can address Shubik (1973)'s critique.\par

The Cournot-Walras equilibrium concept was introduced by Gabszewicz and Vial (1972) in a production economy and it was recast in exchange economies by Codognato and Gabszewicz (1991) and subsequently by Gabszewicz and Michel (1997). These contributions consider exchange economies characterised by few oligopolists (sellers in our case) and many small agents (buyers in our case). Furthermore, following the Cournotian spirit, the oligopolists are allowed to act strategically as they are few, while small agents are assumed to act competitively as price takers as they are many. However this kind of assumption was criticised by Okuno, Postlewaite and Roberts (1980):
\begin{quotation}``Traditional general equilibrium treatments of such situations [in which some but not all agents have market power] have been deficient in that they have simply assumed \textit{a priori} that certain agents behave as price takers while others act non-competitively, with no formal explanation being given as to why a particular agent should behave one way or the other."\end{quotation}

Our Example 2 supports this view as it shows that it is possible to use the Cournot-Walras equilibrium in odd cases where buyers are assumed to behave competitively even if there are only two of them. In other words, the Cournot-Walras approach does not endogenously derive agents' behaviours but it assumes \textit{a priori} that some agents can influence prices while others are price takers.
Another shortcoming of the Cournot-Walras approach is that it is extremely challenging to obtain a general existence result because oligopolists influence prices by manipulating the Walras price correspondence which may fail to be continuous unless strict assumptions are imposed on the set of price-taking agents. Nevertheless, some existence results have been obtained in particular frameworks: Bonisseau and Florig (2003) prove the existence of a Cournot-Walras equilibrium in linear exchange economies; and Codognato and Julien (2013) proved the existence in mixed exchange economies where agents on the continuum have Cobb-Douglas utility functions.\footnote{Shirai (2010) prove the existence of a Cournot-Walras equilibrium in production economies.}

The issue of asymmetrically imposing behavioural assumptions on some agents is overcome in bilateral oligopoly models in which both sellers and buyers are treated symmetrically in that all agents are allowed to behave strategically---no price-taking assumptions are imposed \emph{a priori}---and departing from the Walrasian tradition the price is constructed from agents strategic decisions.

\section{Simultaneous bilateral oligopoly}
As stressed in the introduction, the bilateral oligopoly model is a strategic market game based on a two-commodity exchange economy with corner initial endowments.\footnote{Bloch and Ferrer (2001a) and Dickson and Hartley (2013b) also consider the case in which agents are endowed with both commodities, i.e. have ``interior'' initial endowments, and can choose whether they become sellers or buyers.} We define the exchange economy as $\mathcal{E}=\{(u_i(\cdot),(x_i^0,y_i^0))_{i=1}^{m+n}\}$ as in the previous section. 
The strategy set of a seller $i$ is defined as in (\ref{CE:sellerstrategy}). The strategy set of a buyer $i$ is
$$\mathcal{B}_i=\{b_i\in \mathbb{R}:0\leq b_i\leq y_i^0\}.$$ $b^i$ is the bid of money that buyer $i$ makes on the consumption good. Let $\mathcal{B}=\prod _{i=1}^{m+n}\mathcal{B}_i$ and $\mathcal{B}_{-h}=\prod _{i\neq h}\mathcal{B}_i$. Let $b=(b_{m+1},\dots,b_{m+n})$ and $b_{-i}=(b_{m+1},\dots,b_{i-1},b_{i+1},\dots,b_{m+n})$ be elements of $\mathcal{B}$ and $\mathcal{B}_{-i}$ respectively. For each $(q,b)\in \mathcal{Q}\times\mathcal{B}$, the price vector $p(q,b)=(p_x(q,b),1)$ is determined such that the price of the consumption good is given by the ratio of the total money bids made for the good to the total amount of good made available by the sellers:
\begin{equation}\label{CNE:price}
p_x(q,b) = \left\{ \begin{array}{ll}
        \frac{B}{Q}&\mbox{ if }Q\neq 0 \\
        0&\mbox{ if }Q=0\end{array}\right.,
\end{equation}
with $B=\sum_{i=m+1}^n b^i$. By having defined the price formation rule, which is non-Walrasian, we can now define the rules to calculate the commodity bundles of each agent. For each $(q,b)\in \mathcal{Q}\times\mathcal{B}$, the commodity bundle $(x_i(q,b),y_i(q,b))$ of each seller $i=1,\dots,m$ is given by
\begin{equation}\label{CNE:allocation}\begin{aligned}
x_i(q,b)&=x_i^0-q_i,\\
y_i(q,b)&=p_x(q,b)\cdot q_i,\end{aligned}\end{equation}
and the commodity bundle $(x_i(q,b),y_i(q,b))$ of each buyer $i=m+1,\dots,m+n$ is given by
\begin{align*}
x_i(q,b)&=\frac{b_i}{p_x(q,b)},\\
y_i(q,b)&=y_i^0-b_i.\end{align*}
As such, each agent's amount of the commodity they are endowed with reduces by the quantity of that commodity they offer to the market, and the amount of the other commodity is given by their proportional share of the aggregate amount of that commodity offered by the other side of the market, so sellers receive a share $q_i/Q$ of the aggregate bid $B$, and buyers receive a share $b_i/B$ or the aggregate offer $Q$. The payoff function of agent $i$ is $\pi_i(q,b)=u_i(x_i(q,b),y_i(q,b))$, for $i=1,\dots,m+n$. The simultaneous bilateral oligopoly model is then a set $\Gamma'=\{(\mathcal{Q}_i,\pi_i(\cdot))_{i=1}^{m},(\mathcal{B}_i,\pi_i(\cdot))_{i=m+1}^{m+n}\}$.\par

The definition of a Cournot-Nash equilibrium is as follows.
\begin{definition}The strategy profile $(\hat{q},\hat{b})$ is a Cournot-Nash equilibrium for the game $\Gamma'$ if for each seller $i=1,\ldots,m$ we have $\pi_i(\hat q_i,\hat{q}_{-i},\hat{b})\geq\pi_i(q_i,\hat{q}_{-i},\hat{b})$, for each $q_i\in \mathcal Q_i$, and for each buyer $i=m+1,\ldots,m+n$ we have $\pi_i(\hat{q},\hat b_i,\hat{b}_{-i})\geq\pi_i(\hat{q},b_i,\hat{b}_{-i})$, for each $b_i\in \mathcal B_i$.\end{definition}

We now consider an example to illustrate the Cournot-Nash equilibrium concept.
\begin{example} Consider the exchange economy defined in Example 2. To find the Cournot-Nash equilibrium, we have to solve the payoff maximization problems for all agents. We first consider sellers and then buyers. Consider the maximisation problem of seller 1:
\begin{equation*}\begin{aligned}
&\underset{q_1}{\text{max}}& & \ln(1+(3-q_1))+\frac{b_3+b_4}{q_1+q_2}q_1,\\
& \text{subject to}& & 0\leq q_1\leq 3.\end{aligned}\end{equation*}
This payoff function is concave in $q_1$; we solve as an unconstrained problem then check the solution satisfies the constraints. The first-order condition is
\begin{equation*}\frac{\partial \pi_1}{\partial q_1}:-\frac{1}{4-q_1}+\frac{b_3+b_4}{q_1+q_2}\biggl(1-\frac{q_1}{q_1+q_2}\biggr)=0.\end{equation*}
Since all sellers are identical and all buyers are identical, we consider a symmetric Cournot-Nash equilibrium where $\hat{q}_1=\hat{q}_2$ and $\hat{b}_3=\hat{b}_4$. Then the previous equation becomes,
$$\frac{\partial \pi_1}{\partial q_1}:-\frac{1}{4-q_1}+\frac{\hat{b}_3}{\hat{q}_1}\frac{1}{2}=0,$$
and we obtain
\begin{equation}\label{CNE:q}
\hat{q}_1=\frac{4\hat{b}_3}{2+\hat{b}_3}.
\end{equation}
Consider next the maximisation problem of buyer 3:
\begin{equation*}\begin{aligned}
&\underset{b_3}{\text{max}}& & 3b_3\frac{q_1+q_2}{b_3+b_4}-\frac{1}{2}\biggl(b_3\frac{q_1+q_2}{b_3+b_4}\biggr)^2+(5-b_3),\\
& \text{subject to}& & 0\leq b_3\leq 5.\end{aligned}\end{equation*}
Again this is concave in own strategy and we solve by considering the unconstrained maximisation problem. The first-order condition is
\begin{equation*}\frac{\partial \pi_3}{\partial b_3}:\frac{q_1+q_2}{b_3+b_4}\biggl(1-\frac{b_3}{b_3+b_4}\biggr)\biggl(3-b_3\frac{q_1+q_2}{b_3+b_4}\biggr)-1=0.\end{equation*}
Since we consider the symmetric Cournot-Nash equilibrium, we then obtain
\begin{equation}\label{CNE:b}
\hat{b}_3=\hat{q}_1\frac{1}{2}(3-\hat{q}_1).\end{equation}
By combining equations (\ref{CNE:q}) and (\ref{CNE:b}), we find that the Cournot-Nash equilibrium for the bilateral oligopoly model is
$$(\hat{q}_1,\hat{q}_2,\hat{b}_3,\hat{b}_4)=\biggr(\frac{7-\sqrt{17}}{2},\frac{7-\sqrt{17}}{2},\sqrt{17}-3,\sqrt{17}-3\biggl).$$
At the Cournot-Nash equilibrium the price vector is $p(\hat{q},\hat{b})=(\frac{\sqrt{17}-1}{4},1)$ and the allocation is
\begin{align*}
&(x_i(\hat{q},\hat{b}),y_i(\hat{q},\hat{b}))=\biggl(\frac{\sqrt{17}-1}{2},\sqrt{17}-3\biggr),\mbox{ for }i=1,2,\\
&(x_i(\hat{q},\hat{b}),y_i(\hat{q},\hat{b}))=\biggl(\frac{7-\sqrt{17}}{2},8-\sqrt{17}\biggr),\mbox{ for }i=3,4.
\end{align*}\qed\end{example}

This example clarifies that the simultaneous bilateral oligopoly model allows us to study oligopoly where all agents are treated symmetrically and no assumptions on their behaviours are made \textit{a priori}. Indeed, given the structure of the game, all agents act strategically because they can manipulate the price by changing their actions. Therefore bilateral oligopoly addresses the critique raised by Okuno et al. (1980). The fact that buyers act strategically is one of the reasons why sellers' and buyers' commodity bundles at the Cournot-Nash equilibrium are different from those obtained at the Cournot-Walras equilibrium, even if the exchange economies in Examples 2 and 3 are the same. We stress again that buyers are assumed to behave as price-takers in the Cournot-Walras approach while they are permitted to act strategically in bilateral oligopoly. Furthermore, while the Cournot-Walras model suffers from issues because of the potential discontinuity of the Walras price correspondence, this is not the case in bilateral oligopoly because the price formation rule is non-Walrasian. The existence of a Cournot-Nash equilibrium was studied by Bloch and Ferrer (2001a) and Dickson and Hartley (2008), who also studied uniqueness of equilibrium. Additionally, Bloch and Ferrer (2001b) showed the existence when all agents have a constant elasticity of substitution utility function.\par

At this point of the analysis a natural question arises: is it possible to employ the bilateral oligopoly model to study asymmetric oligopoly where no \textit{a priori} assumptions on agents behaviours are made? In other words, is it possible to develop a framework where the differences in agents' behaviours arise endogenously in equilibrium? This is also known as the problem of providing a strategic foundation for oligopoly. 

In the literature, two main approaches are used to this end. The first approach consists in using mixed exchange economies where large agents (oligopolists) are represented by atoms and small agents (competitive agents) by an atomless continuum. An atom is an agent whose initial endowment is non-negligible compared to the total endowment of the economy while a agent in the continuum holds only a negligible part of it. The model was proposed by Gabszewicz and Mertens (1971) and Shitovitz (1973) in cooperative game theory. Shitovitz (1973) stated:
\begin{quotation}``The main point in our treatment is that the small and the large traders are not segregated into different groups a priori; they are treated on exactly the same basis. The distinctions we have found between them are an outcome of the analysis; they have not been artificially introduced in the beginning, as is the case in the classical approach."\end{quotation} This approach based on mixed exchange economies was also extended to noncooperative game theory by Okuno et al. (1980), in order to address their critique, and further generalised by Busetto, Codognato, and Ghosal (2011) and Busetto, Codognato, Ghosal, Julien, and Tonin (2017). In this setting, while all agents have \textit{a priori} the same strategic position, they found that in equilibrium large agents represented by atoms have market power while small agents on the continuum behave as if they are price-takers.\par

The second approach is based on considering partial replicas of exchange economies where the number of some agents is increased while proportionally reducing their weights. This method was introduce by Dr\`eze, Gabszewicz, and Gepts (1969) in cooperative game theory and it was adapted to bilateral oligopoly by Dickson and Hartley (2013). The main result of this approach is that agents who are replicated lose their ability to influence the prices and at the limit they behave competitively whereas the others keep market power.\par

In the last part of the section, we consider this second approach and we study the limit of a partial replica where we replicate only the set of buyers. We study such limit  because it represents an asymmetric oligopoly where buyers are asymptotically negligible and so should behave as price-takers. Consider an exchange economy $\mathcal{E}=\{(u_i(\cdot),(x_i^0,y_i^0))_{i=1}^{m+n}\}$ where the first $m$ agents are sellers and all other $n$ agents are buyers. We associate to each buyer a weight $\frac{1}{n}$ which is used to scale their initial endowment when compared to the whole economy. Therefore, the total initial endowment of money is give by $\sum_{i=m+1}^{m+n}\frac{1}{n}y_i^0$. When $n$ increases, the number of buyers increases but the total initial endowment of money remains the same as in the initial economy. Given the price formation rule (\ref{CNE:price}) and since $B=\sum_{i=m+1}^{m+n}\frac{1}{n}b_i$, if $n$ increases, then the buyers' ability to influence prices decreases and at the limit, for $n\rightarrow\infty$, totally disappears. It is important to stress that the limit of a sequence of Cournot-Nash equilibria, which is denoted by $(\bar{q},\bar{b})$, is not a Cournot-Nash equilibrium of the limit economy as the underlying game in the limit is not well-defined. Nevertheless, the limit of the sequence gives us an object that can be compared to the other solution concepts in finite economies.\par

We now consider an example to illustrate how to partially replicate an exchange economy and how to calculate its limit.\par
\begin{example}Consider an exchange economy with 2 sellers and $2n$ buyers such that $u_i(x,y)=\ln (1+x)+y$ and $(x_i^0,y_i^0)=(3,0)$, for $i=1,2$, and $u_i(x,y)=3x-\frac{1}{2}x^2+y$ and $(x_i^0,y_i^0)=(0,5)$, for $i=3,\dots,2n+2$. As such, it is as if there are $n$ copies of both the buyers 3 and 4 defined in Example 2. To each buyer is associated a weight $\frac{1}{n}$ and then $B=\sum_{i=3}^{2n+2}\frac{1}{n}b_i$. To simplify the analysis, we consider a symmetric Cournot-Nash equilibrium, where all buyers play the same strategy, and then $\hat{B}=2\hat{b}_i$. But then, if we consider the sellers' payoff maximisation problems, it is straightforward to verify that sellers' best responses are as in the previous example
\begin{equation}\label{CNER:q}
\hat{q}_i=\frac{4\hat{b}_3}{2+\hat{b}_3},
\end{equation}
for $i=1,2$. On the contrary, buyers' maximisation problems are different from the previous example because now there are $2n$ buyers instead of two. Consider the maximisation problem of a typical buyer 3:
\begin{equation}\label{CNER:maxbuyer}\begin{aligned}
&\underset{b_3}{\text{max}}& & 3b_3\frac{q_1+q_2}{\frac{1}{n}b_3+\sum_{i=4}^{2n+2}\frac{1}{n}b_i}-\frac{1}{2}\biggl(b_3\frac{q_1+q_2}{\frac{1}{n}b_3+\sum_{i=4}^{2n+2}\frac{1}{n}b_i}\biggr)^2+(5-b_3),\\
& \text{subject to}& & 0\leq b_3\leq 5.
\end{aligned}\end{equation}
To simplify the notation, let $B_{-3}=\sum_{i=4}^{2n+2}\frac{1}{n}b_i$. The first order condition of the buyer's unconstrained maximisation problem can be written as
\begin{equation}\label{CNER:focb}\frac{\partial \pi_3}{\partial b_3}:\frac{q_1+q_2}{\frac{1}{n}b_3+B_{-3}}\biggl(1-\frac{\frac{1}{n}b_3}{\frac{1}{n}b_3+B_{-3}}\biggr)\biggl(3-b_3\frac{q_1+q_2}{\frac{1}{n}b_3+B_{-3}}\biggr)-1=0.\end{equation}
Since we consider a symmetric Cournot-Nash equilibrium where $\hat{q}_1=\hat{q}_2$ and $\hat{b}_3=\dots=\hat{b}_{2n+2}$, from the previous equation we obtain
\begin{equation}\label{CNER:b}\hat{b}_3=\hat{q}_1\biggl(1-\frac{1}{2n}\biggr)(3-\hat{q}_1).\end{equation}
Let $k=1-\frac{1}{2n}$. By combining equations (\ref{CNER:q}) and (\ref{CNER:b}), we obtain the following Cournot-Nash equilibrium for the bilateral oligopoly model
\begin{equation*}\begin{split}(\hat{q}_1,\hat{q}_2,\hat{b}_3,\dots,\hat{b}_{2n+2})=\biggr(\frac{7 k-\sqrt{k^2+8k}}{2k},\frac{7 k-\sqrt{k^2+8k}}{2k},\\2\sqrt{k^2+8k}-2k-2,\dots,2\sqrt{k^2+8k}-2k-2\biggl)\end{split}\end{equation*}
The limit of this Cournot-Nash equilibrium for $n\rightarrow\infty$ is $\bar{q}_i=2$ for sellers and $\bar{b}_i=2$ for buyers. At the limit the price vector is $p(\bar{q},\bar{b})=(1,1)$  and the sellers' and buyers' commodity bundles are
\begin{align*}
&(x_i(\bar{q},\bar{b}),y_i(\bar{q},\bar{b}))=(1,2),\mbox{ for a seller }i,\\
&(x_i(\bar{q},\bar{b}),y_i(\bar{q},\bar{b}))=(2,3),\mbox{ for a buyer }i.\end{align*}
\qed\end{example}
It is immediate to see that the commodity bundles at the limit of the Cournot-Nash equilibrium are different from the ones at the Cournot-Walras equilibrium of Example 2. This result is somewhat surprising as at the limit of the Cournot-Nash equilibrium buyers have no influence on the price, as is assumed in the Cournot-Walras approach, yet the equilibrium allocations differ. This is due to the fact that the Cournot-Walras approach has an intrinsic two-stage structure---sellers decide their supply in the first stage taking as given the Walrasian demands of the buyers---whereas when considering a Cournot-Nash equilibrium in the bilateral oligopoly model all agents choose their actions simultaneously so sellers must form beliefs about how the buyers will behave. This fact was stressed by Dickson and Hartley (2013a) who considered the conditions under which the two models coincide in the limit, and when they do not. In mixed exchange economies, Codognato (1995) and Busetto, Codognato, and Ghosal (2008) showed that the allocation at a Cournot-Walras equilibrium is different from the one at a Cournot-Nash equilibrium of the Shapley window model. Therefore, we can conclude that the bilateral oligopoly model can be useful to study simultaneous asymmetric oligopoly in a closed market model by partially replicating the underlying exchange economy but it does not provide a foundation of the Cournot-Walras approach. In the next section, we address this last issue by studying a sequential bilateral oligopoly model.

\section{Sequential bilateral oligopoly}
In this section we consider a sequential-move bilateral oligopoly model with a two-stage structure where the timing of the model is exogenously specified as follows: in the first stage sellers simultaneously choose the quantities of the consumption good to put up in exchange for money; in the second stage buyers simultaneously choose the quantity of money to bid for the consumption good. At the end of the second stage bids and offers are aggregated and price is determined according to the usual rule in (\ref{CNE:price}). Groh (1999) first considered a sequential reformulation of bilateral oligopoly in the context of an example, that was extended to general settings in Dickson (2006).\footnote{Dickson and Hartley (2013) consider a sequential market share game and Busetto et al. (2008) consider a sequential reformulation of the Shapley window model.}\par

We first consider an exchange economy $\mathcal{E}=\{(u_i(\cdot),(x_i^0,y_i^0))_{i=1}^{m+n}\}$ and we then define the sequential bilateral oligopoly. The strategy set of a seller $i$ is defined as in (\ref{CE:sellerstrategy}). The strategy set of a buyer $i$ is
$$\mathcal{B}_i=\{b_i(\cdot) \mbox{ is a function such that } b_i:\mathcal{Q}\rightarrow[0,y_i^0]\}.$$
Let $\mathcal{B}=\prod _{i=1}^{m+n}\mathcal{B}_i$ and $b(\cdot)=(b_{m+1}(\cdot),\dots,b_{m+n}(\cdot))$ be an element of $\mathcal{B}$.
For each $(q,b(q))\in \mathcal{Q}\times\mathcal{B}$, the price vector $p(q,b(q))=(p_x(q,b(q)),1)$ is such that
\begin{equation}\label{SPNE:price}
p_x(q,b(q)) = \left\{ \begin{array}{ll}
        \frac{B(q)}{Q}&\mbox{ if }Q\neq 0 \\
        0&\mbox{ if }Q=0\end{array}\right.,
\end{equation}
with $B(q)=\sum_{i=m+1}^n b_i(q) $. For each $(q,b(q))\in \mathcal{Q}\times\mathcal{B}$, the commodity bundle $(x_i(q,b(q)),y_i(q,b(q)))$ of a seller $i$ is given by
\begin{align*}
x_i(q,b(q))&=x_i^0-q_i,\\
y_i(q,b(q))&=p_x(q,b(q))\cdot q_i,\end{align*}
for $i=1,\dots,m$, and the commodity bundle $(x_i(q,b(q)),y_i(q,b(q)))$ of a buyer $i$ is given by
\begin{align*}
x_i(q,b(q))&=\frac{b_i(q)}{p_x(q,b(q))},\\
y_i(q,b(q))&=y_i^0-b_i(q),\end{align*}
for $i=m+1,\dots,m+n$.
The payoff function of an agent $i$ is $\pi_i(q,b(q))=u_i(x_i(q,b(q)),$ $y_i(q,b(q)))$, for $i=1,\dots,m+n$. The sequential bilateral oligopoly model is then a set $\Gamma''=\{(\mathcal{Q}_i,\pi_i(\cdot))_{i=1}^{m},(\mathcal{B}_i,\pi_i(\cdot))_{i=m+1}^n\}$.\par
We now define a subgame perfect Nash equilibrium (hereafter SPNE) which is the equilibrium concept we use in this dynamic game.
\begin{definition}A strategy profile $(\hat{q},\hat{b})$ is a SPNE for $\Gamma''$ if an only if it is a Cournot-Nash equilibrium in every subgame of the game $\Gamma''$.\end{definition}
Note that in the sequential bilateral oligopoly the subgames are the whole game and the subgame in which buyers choose their optimal bids following any vector of offers from the sellers. We now consider an example to illustrate how to find the SPNE and how to find the limit of a SPNE when we partially replicate the exchange economy. We stress again that the limit is not a SPNE as the underlying game is not well-defined at the limit.
\begin{example}Consider the same exchange economy defined in Example 4. As before, to simplify the analysis, we consider a symmetric SPNE that can be computed as follows. We first calculate the optimal buyers' strategy for any feasible strategies of sellers. This guarantees to find a Cournot-Nash equilibrium in the buyers' subgame. We then substitute the optimal buyers' strategy in the sellers' payoff functions to determine their reaction functions given the responses of the buyers, after which we can find mutually consistent best responses. The strategy profile obtained in such way is a SPNE. Consider a typical buyer 3. It is immediate to verify that the buyer's maximisation problem is as in (\ref{CNER:maxbuyer}). Therefore the best response of buyer 3 solves the first order condition in (\ref{CNER:focb}). Since we consider a symmetric SPNE where all buyers play the same strategy, equation (\ref{CNER:focb}) can be rewritten as
\begin{equation*}\frac{\partial \pi_3}{\partial b_3}:\frac{q_1+q_2}{2b_3(q)}\biggl(1-\frac{1}{2n}\biggr)\biggl(3-\frac{q_1+q_2}{2}\biggr)-1=0.\end{equation*}
We then obtain
\begin{equation}\label{SPNE:b}b_3(q)=\frac{q_1+q_2}{2}k\biggl(3-\frac{q_1+q_2}{2}\biggr),\end{equation}
with $k=1-\frac{1}{2n}$. Consider next the maximisation problem of seller 1. If we substitute the strategy $b_3(q)$ in the seller's payoff function we obtain
\begin{equation}\label{SPNE:max}\begin{aligned}
&\underset{q_1}{\text{max}}& & \ln(1+(3-q_1))+2\frac{q_1+q_2}{2}k\biggl(3-\frac{q_1+q_2}{2}\biggr)\frac{1}{q_1+q_2}q_1,\\
& \text{subject to}& & 0\leq q_1\leq 3.\end{aligned}\end{equation}
The first order condition of the seller's unconstrained maximisation problem is
\begin{equation*}\frac{\partial \pi_1}{\partial q_1}:-\frac{1}{4-q_1}+\frac{6k-2kq_1-kq_2}{2}=0.\end{equation*}
Since we consider a symmetric SPNE where $\hat{q}_1=\hat{q}_2$, the previous equation becomes
\begin{equation}\label{SPNE:q}q_1=\frac{9k-\sqrt{9k^2+6 k}}{3k}.\end{equation}
Therefore, by combining equations (\ref{SPNE:b}) and (\ref{SPNE:q}), we obtain the following SPNE for the sequential bilateral oligopoly
\begin{equation*}\begin{split}(\hat{q}_1,\hat{q}_2,\hat{b}_3(\hat{q}),\dots,\hat{b}_{2n+2}(\hat{q}))=\biggr(\frac{9k-\sqrt{9k^2+6k}}{3k},\frac{9k-\sqrt{9k^2+6k}}{3k},\\ \frac{3\sqrt{9k^2+6k}-3k-2}{3},\dots,\frac{3\sqrt{9k^2+6k}-3k-2}{3}\biggl).\end{split}\end{equation*}
The limit of the SPNE for $n\rightarrow\infty$ is $\bar{q}_i=\frac{9-\sqrt{15}}{3}$ for sellers and $\bar{b}_i(\bar{q})=\frac{3\sqrt{15}-5}{3}$ for buyers. At the limit the price vector is $p(\bar{q},\bar{b}(\bar{q}))=(\frac{\sqrt{17}-1}{4},1)$ and the sellers' and buyers commodity bundles are
\begin{align*}
&(x_i(\bar{q},\bar{b}(\bar{q})),y_i(\bar{q},\bar{b}(\bar{q})))=\biggl(\frac{\sqrt{17}-1}{2},\sqrt{17}-3\biggr),\mbox{ for a seller }i,\\
&(x_i(\bar{q},\bar{b}(\bar{q})),y_i(\bar{q},\bar{b}(\bar{q}))=\biggl(\frac{7-\sqrt{17}}{2},8-\sqrt{17}\biggr),\mbox{ for a buyer }i.\end{align*}
\qed\end{example}
It is immediate to see that the commodity bundles at the limit of the SPNE are different to those at the limit of the Cournot-Nash equilibrium in which moves are simultaneous, but are the same as those at the Cournot-Walras equilibrium. We stress again that in this framework all agents are treated symmetrically and the different behaviours of sellers and buyers are obtained endogenously at the limit of the SPNE. With this last example, we have shown that sequential bilateral oligopoly in which it is exogenously specified that sellers move first and buyers move second can provide a closed market model to study sequential asymmetric oligopoly by partially replicating the underlying exchange economy. Furthermore this model can provide a foundation for the Cournot-Walras approach. It is worth stressing that the price equation (\ref{SPNE:price}) in the limit, when $n\rightarrow\infty$, corresponds to the inverse demand function in Example 1. In mixed exchange economies, Busetto et al. (2008) showed that the set of the Cournot-Walras equilibrium allocations coincides with a particular subset of SPNE allocations of the two-stage reformulation of the Shapley window model.

\section{Walras equilibrium and welfare}
The Walrasian analysis in the synthesis reached in the contributions of Debreu (1959), Arrow and Hahn (1971) and Mckenzie (2002) crucially relies on the price-taking assumption, i.e., all agents are assumed to behaved competitively. It is then important to study under which conditions on the fundamentals of an economy agents consider prices as given endogenously in equilibrium, without making \textit{ad hoc} assumptions. This is the problem of providing a strategic foundation for perfect competition, which is similar in spirit to the problem of finding a foundation for models of oligopoly. In fact, a strategic foundation for perfect competition can also be provided by considering continuum exchange economies or replicated exchange economies.\footnote{These different approaches are illustrated by the distinction between limit theorems and theorems in the limit (see Gale 2000). See Mas-Colell (1982) for a study on the links between the two approaches.} In cooperative game theory Aumann (1964) provided a foundation for perfect competition in economies with a continuum of agents by showing an equivalence result between the core and the Walras allocation. Subsequently, in noncooperative game theory, Dubey and Shapley (1994) and Codognato and Ghosal (2000) considered strategic market games with a continuum of agents and they show equivalence results between the Cournot-Nash and Walras allocations. Those results are based on the fact that when there is a continuum of agents everyone is negligible and cannot influence prices. The second approach based on replicated exchange economies was pioneered in cooperative game theory by Edgeworth (1881) and further analysed by Debreu and Scarf (1963). Subsequently, Dubey and Shubik (1978), Sahi and Yao (1989), and Amir et al. (1990) applied this technique to strategic market games. They show that when all agents in the economy are replicated the allocation at the limit of the Cournot-Nash equilibrium corresponds to the allocation at the Walras equilibrium.\footnote{Note that when all agents in the economy are replicated it is not necessary to decrease their weight.}  Those results are heuristically based on the fact that when the number of agents increases their influence on prices decreases and at the limit totally disappears.\footnote{Mas-Colell (1980) surveyed the main contributions on the foundations of perfect competition by considering other noncooperative approaches that do not rely on strategic market games.}\par

We now show how the bilateral oligopoly model can be employed to provide a foundation for perfect competition. Let $\mathcal{E}=\{(u_i(\cdot),(x_i^0,y_i^0))_{i=1}^{m+n}\}$. We first define a Walras equilibrium concept.
\begin{definition} A Walras equilibrium is pair $(p^*,(x_i^*,y_i^*)_{i=1}^{n+m})$ of a price vector and an allocation such that each commodity bundle $(x_i^*,y_i^*)$ maximises agent $i$'s utility function within his budget set, for $i=1,\dots,m+n$, and all markets clear, i.e., $\sum_{i=1}^{m+n} x_i^*=\sum_{i=1}^{m+n}x_i^0$ and $\sum_{i=1}^{m+n} y_i^*=\sum_{i=1}^{m+n}y_i^0$.\end{definition}
From the analysis made in Section 4, we can conjecture that if both the number of sellers and buyers are replicated while proportionally reducing their weights, all agents lose market power and the sellers' and buyers' commodity bundles at the limit of the Cournot-Nash equilibrium will correspond to ones at the Walras equilibrium. We now consider an example to illustrate this point.
\begin{example}Consider an exchange economy with $2n$ sellers and $2n$ buyers such that $u_i(x,y)=\ln (1+x)+y$ and $(x_i^0,y_i^0)=(3,0)$, for $i=1,\dots,2n$, and $u_i(x,y)=3x-\frac{1}{2}x^2+y$ and $(x_i^0,y_i^0)=(0,5)$, for $i=2n+1,\dots,4n$. It is basically as the exchange economy defined in Example 2 with $n$ copies of both sellers 1 and 2 and buyers 3 and 4. The Walras equilibrium of the exchange economy is\footnote{It is straightforward to verify that sellers' and buyers' commodity bundles are the same for any $n$.}
\begin{align*}&(p^{1*},p^{2*})=\biggl(\frac{\sqrt{5}-1}{2},1\biggr),\\
&(x_i^{1*},x_i^{2*})=\biggl(\frac{\sqrt{5}-1}{2},2\sqrt{5}-3\biggr),\mbox{ for }i=1,\dots,2n\\
&(x_i^{1*},x_i^{2*})=\biggl(\frac{7-\sqrt{5}}{2},8-2\sqrt{5}\biggr),\mbox{ for }i=2n+1,\dots,4n.\end{align*}
We now compute the Cournot-Nash equilibrium. As before, we first consider the maximisation problem of seller $1$.
\begin{equation*}\begin{aligned}
&\underset{q_1}{\text{max}}& & \ln(1+(3-q_1))+\frac{\sum_{i=2n+1}^{4n} \frac{1}{n}b_i}{\frac{1}{n}q_1+\sum_{i=2}^{2n}\frac{1}{n}q_i}q_1,\\
& \text{subject to}& & 0\leq q_1\leq 3.\end{aligned}\end{equation*}
The first order condition of the seller's unconstrained maximisation problem is
\begin{equation*}\frac{\partial \pi_1}{\partial q_1}:-\frac{1}{4-q_1}+\frac{\sum_{i=2n+1}^{4n}\frac{1}{n}b_i}{\frac{1}{n}q_1+\sum_{i=2}^{2n}\frac{1}{n}q_i}\biggl(1-\frac{q_1}{\frac{1}{n}q_1+\sum_{i=2}^{2n}\frac{1}{n}q_i}\biggr)=0.\end{equation*}
Since we consider a symmetric Cournot-Nash equilibrium where all sellers play the same strategy, we obtain
\begin{equation}\label{WE:q}\hat{q}_1=\frac{4\hat{b}_{2n+1}k}{1+\hat{b}_{2n+1}k},\end{equation}
with $k=1-\frac{1}{2n}$.\par
Consider now the maximisation problem of buyer $2n+1$.
\begin{equation*}\begin{aligned}
&\underset{b_3}{\text{max}}& & 3b_{2n+1}\frac{\sum_{i=1}^{2n}\frac{1}{n}q_i}{\frac{1}{n}b_{2n+1}+\sum_{i=2n+2}^{4n}\frac{1}{n}b_i}-\frac{1}{2}\biggl(b_{2n+1}\frac{\sum_{i=1}^{2n}\frac{1}{n}q_i}{\frac{1}{n}b_{2n+1}+\sum_{i=2n+2}^{4n}\frac{1}{n}b_i}\biggr)^2+(5-b_{2n+1}),\\
& \text{subject to}& & 0\leq b_{2n+1}\leq 5.
\end{aligned}\end{equation*}
The first order condition of the buyer's unconstrained maximisation problem is
\begin{equation*}\begin{split}\frac{\partial \pi_{2n+1}}{\partial b_{2n+1}}:\frac{\sum_{i=1}^{2n}\frac{1}{n}q_i}{\frac{1}{n}b_{2n+1}+\sum_{i=2n+2}^{4n}\frac{1}{n}b_i}\biggl(1-\frac{b_{2n+1}}{\frac{1}{n}b_{2n+1}+\sum_{i=2n+2}^{4n}\frac{1}{n}b_i}\biggr)\\ \biggl(3-b_{2n+1}\frac{\sum_{i=1}^{2n}\frac{1}{n}q_i}{\frac{1}{n}b_{2n+1}+\sum_{i=2n+2}^{4n}\frac{1}{n}b_i}\biggr)-1=0.\end{split}\end{equation*}
Since we consider the symmetric Cournot-Nash equilibrium where all buyers play the same strategy, we finally obtain
\begin{equation}\label{WE:b}
\hat{b}_{2n+1}=\hat{q}_1k(3-\hat{q}_1).\end{equation}
By combining equations (\ref{WE:q}) and (\ref{WE:b}), we obtain the following Cournot-Nash equilibrium for the bilateral oligopoly model
\begin{equation*}\begin{split}(\hat{q}_1,\dots,\hat{q}_{2n},\hat{b}_{2n+1},\dots,b_{4n})=\biggr(\frac{7k-\sqrt{k^2+4}}{2k},\dots,\frac{7k-\sqrt{k^2+4}}{2k},\\\frac{2k\sqrt{k^2+4}-2k^2-1}{k},\dots,\frac{2k\sqrt{k^2+4}-2k^2-1}{k}\biggl).\end{split}\end{equation*}
The limit of this Cournot-Nash equilibrium as $n\rightarrow\infty$ is $\bar{q}_i=\frac{7-\sqrt{5}}{2}$ for sellers and $\bar{b}_i=2\sqrt{5}-3$ for buyers. At the limit the price vector is $p(\bar{q},\bar{b})=(\frac{\sqrt{5}-1}{2},1)$ and the sellers' and buyers' commodity bundles are
\begin{align*}
&(x_i(\bar{q},\bar{b}),y_i(\bar{q},\bar{b}))=\biggl(\frac{\sqrt{5}-1}{2},2\sqrt{5}-3\biggr),\mbox{ for a seller }i,\\
&(x_i(\bar{q},\bar{b}),y_i(\bar{q},\bar{b}))=\biggl(\frac{7-\sqrt{5}}{2},8-2\sqrt{5}\biggr),\mbox{ for a buyer }i.\end{align*}\qed\end{example}
It is immediate to see that the commodity bundles at the limit of the Cournot-Nash equilibrium correspond to the ones at the Walras equilibrium of the underlying exchange economy. This example further clarifies that the assumption of competitive behaviour is justified when there are many of each type of agent.\par

It is worth noting that this convergence to the Walras equilibrium in terms of commodity bundles, when all agents are replicated, is also obtained with the Cournot-Walras equilibrium and the SPNE. These results, obtained for particular examples, suggest that the timing of the model is not relevant when considering a foundation for perfect competition. Lahmandi-Ayed (2001) proved the converge of a Cournot-Walras equilibrium to the Walras equilibrium in exchange economies. As mentioned above, Dubey and Shubik (1978), Amir et al. (1990), and Sahi and Yao (1989) studied the converge of a Cournot-Nash equilibrium to the Walras equilibrium in strategic market games. Koutsougeras and Meo (2017) considered the convergence of a Cournot-Nash equilibrium to the Walras equilibrium for general sequences of economies whose distribution of characteristics has compact support by using the model defined by Peck et al. (1992).\par

We conclude this section by making some welfare considerations on the equilibrium concepts used throughout the paper. First of all, by the First Welfare Theorem, the Walras equilibrium is Pareto efficient. Differently, all other allocations found in Examples 2--5 are Pareto inefficient. In fact, it is possible to find other commodity bundles which Pareto dominate them. In the following table we report the utility levels of sellers and buyers in the different examples.
\begin{small}
\begin{center}\begin{tabular}{lcccc}\hline
&\multicolumn{2}{c}{\textbf{Commodity Bundles}} & \multicolumn{2}{c}{\textbf{Utility levels}}\\ \hline
&\textbf{sellers} & \textbf{buyers} & \textbf{sellers} & \textbf{buyers}\\ \hline
Cournot-Walras (Ex. 2) &$\Bigl(\frac{\sqrt{15}}{3},\frac{3\sqrt{15}-5}{3}\Bigr)$&$\Bigl(\frac{9-\sqrt{15}}{3},\frac{20-3\sqrt{15}}{3}\Bigr)$&3.035&6.460 \\ \hline
Cournot-Nash (Ex. 3) &$\bigl(\frac{\sqrt{17}-1}{2},\sqrt{17}-3\Bigr)$&$\Bigl(\frac{7-\sqrt{17}}{2},8-\sqrt{17}\Bigr)$&2.064&7.158 \\ \hline
Cournot-Nash replica (Ex. 4) &$(1,2)$&$(2,3)$&2.693&7.000 \\ \hline
SPNE replica (Ex. 5) &$\Bigl(\frac{\sqrt{15}}{3},\frac{3\sqrt{15}-5}{3}\Bigr)$&$\Bigl(\frac{9-\sqrt{15}}{3},\frac{20-3\sqrt{15}}{3}\Bigr)$&3.035&6.460 \\ \hline
Walras equilibrium (Ex. 6) &$\Bigl(\frac{\sqrt{5}-1}{2},2\sqrt{5}-3\Bigr)$&$\Bigl(\frac{7-\sqrt{5}}{2},8-2\sqrt{5}\Bigr)$&1.950&7.837 \\ \hline
\end{tabular}\end{center}
\end{small}
It is interesting to note that at the Walras equilibrium, the only efficient equilibrium, sellers get the lowest utility while buyers get the highest. Furthermore, sellers get the highest utility at the Cournot-Walras equilibrium which means that, in our examples, they are better off when they face competitive buyers in a sequential oligopoly. We finally remark that Examples 2--6 belong to the general equilibrium framework while Example 1 is a model of partial analysis. For this reason the efficiency of that model cannot be evaluated by using the notion of Pareto efficiency

\section{Conclusion}
In this paper we have provided a unified framework to compare different equilibrium concepts to study imperfect competition in exchange economies. We have also clarified the differences among symmetric, asymmetric, simultaneous, and sequential oligopoly and analysed the relationships among them in terms of replicated exchange economies and equilibrium concepts.\par

By studying our examples we have found some interesting facts. First, we have seen that the Cournot-Walras equilibrium corresponds to the limit of a SPNE where only buyers are replicated.  Second, we mentioned that when all agents in the exchange economies are replicated the commodity bundles at the limit of any equilibrium concept correspond to the ones at the Walras equilibrium. This suggests that when there is perfect competition the timing of the model is not relevant. We believe that further research should be devoted on those two issues in order to establish more general results.\par

It is important to remark that asymmetric oligopoly can be studied by using the partial replica or by considering mixed exchange economies. This alternative approach was introduced by Okuno et al. (1980) and further scrutinised by Busetto, Codognato, and Ghosal (2008, 2011, 2017) and Busetto, Codognato, Ghosal, Julien, and Tonin (2017). Furthermore, we have focused on exchange economies because economies with production raise many theoretical and technical problems (see Hart (1985), Gary-Bobo (1988), Bonanno (1990) for a survey). To the best of our knowledge, the only contributions on strategic market games with production are Dubey and Shubik (1977) and Chen, Korpeoglu, and Spear (2017). Finally, we have restricted our analysis to exchange economies with only two commodities because in a framework with more commodities the results depend on the particular type of strategic market game adopted. \par

We conclude by mentioning that there is a small literature on experiments in bilateral oligopoly whose contributions are Duffy, Matros, and Temzelides (2011) and Barreda-Tarrazona, García-Gallego, Georgantzís, and Ziros (2015). We expect this line of inquiry to become more popular in the coming years.


\begin{thebibliography}{99}
\bibitem{1}Amir R., Bloch F. (2009). Comparative statics in a simple class of strategic market games. \textit{Games and Economic Behavior} \textbf{65}, 7-24.
\bibitem{2}Amir R., Sahi S., Shubik M., Yao S. (1990). A strategic market game with complete markets. \textit{Journal of Economic Theory} \textbf{51}, 126-143.
\bibitem{3}Arrow K.J., Hahn F.H. (1971).\textit{General competitive analysis}. Holden Day, San Francisco.
\bibitem{4}Aumann R.J. (1964). Markets with a continuum of traders. \textit{Econometrica} \textbf{32}, 39-50.
\bibitem{5}Barreda-Tarrazona I., García-Gallego A., Georgantzís N., Ziros N. (2015). Market games as social dilemmas. \textit{Working Paper} No. 07-2015, University of Cyprus.
\bibitem{6}Bloch F., Ghosal S. (1997). Stable trading structures in bilateral oligopolies. \textit{Journal of Economic Theory} \textbf{74}, 368-384.
\bibitem{7}Bloch F., Ferrer H. (2001a). Trade fragmentation and coordination in strategic market games. \textit{Journal of Economic Theory} \textbf{101}, 301-316.
\bibitem{8}Bloch F., Ferrer H. (2001b). Strategic complements and substitutes in bilateral oligopolies. \textit{Economics Letters} \textbf{70}, 83–87.
\bibitem{9}Bonanno G. (1990). General equilibrium theory with imperfect competition. \textit{Journal of Economic Surveys} \textbf{4}, 297-328.
\bibitem{10}Bonisseau J., Florig M. (2003). Existence and optimality of oligopoly equilibria in linear exchange economies. \textit{Economic Theory} \textbf{22}, 727–741.
\bibitem{11}Busetto F., Codognato G., Ghosal S. (2008). Cournot-Walras equilibrium as a subgame perfect equilibrium. \textit{International Journal of Game Theory} \textbf{37}, 371-386.
\bibitem{12}Busetto F., Codognato G., Ghosal S. (2011). Noncooperative oligopoly in markets with a continuum of traders. \textit{Games and Economic Behavior} \textbf{72}, 38-45.
\bibitem{13}Busetto F., Codognato G., Ghosal S. (2013). Three models of noncoopative oligopoly in markets with a continuum of traders. \textit{Recherches \'Economiques de Louvain}, \textbf{79} 5-13.
\bibitem{14}Busetto F., Codognato G., Ghosal S. (2017). Asymptotic equivalence between Cournot-Nash and Walras equilibria in exchange economies with atoms
and an atomless part. \textit{International Journal of Game Theory}
\bibitem{15}Busetto F., Codognato G., Ghosal S., Julien L., Tonin S. (2017). Noncooperative oligopoly in markets with a continuum of traders and a strongly connected set of commodities. \textit{Games and Economic Behavior}
\bibitem{16}Chen G., Korpeoglu C.G., and Spear S.E. (2017). Price stickiness and markup variations in market games. \textit{Journal of Mathematical Economics}.
\bibitem{17}Codognato G. (1995). Cournot-Walras and Cournot equilibria in mixed markets: a comparison. \textit{Economic Theory} \textbf{5}, 361-370.
\bibitem{18}Codognato G., Gabszewicz J.J. (1991). \'Equilibres de Cournot-Walras dans une économie d'échange. \textit{Revue \'Economique} \textbf{42}, 1013-1026.
\bibitem{57}Codognato G., Ghosal S. (2000). Cournot-Nash equilibria in limit exchange economies with complete markets and consistent prices. \textit{Journal of Mathematical Economics} \textbf{34}, 39-53.
\bibitem{55}Codognato G., Julien L. (2013). Noncooperative oligopoly in markets with
a Cobb-Douglas continuum of traders. \textit{Recherches \'Economiques de Louvain}, \textbf{79} 75-88.
\bibitem{19}Cournot A.A. (1838). \textit{Recherches sur les principes mathématiques de la théorie des richesses}. Hachette, Paris.
\bibitem{20}Debreu G. (1959). \textit{Theory of Value}. Yale University Press, New Haven.
\bibitem{21}Debreu G., Scarf H. (1963). A limit theorem on the core of an economy. \textit{International Economic Review} \textbf{4}, 235-246.
\bibitem{}Dickson A. (2006). The Strategic Marshallian Cross; Shapley and Shubik meet Marshall and Cournot. PhD Thesis, University of Keele.
\bibitem{22}Dickson A. (2013a). The Effects of Entry in Bilateral Oligopoly. \textit{Games} \textbf{4}, 283-303.
\bibitem{23}Dickson A. (2013b). On Cobb-Douglas Preferences in Bilateral Oligopoly.
\textit{Recherches \'economiques de Louvain} \textbf{79}, 89-110.
\bibitem{24}Dickson A., Hartley R. (2008). The strategic Marshallian cross. \textit{Games and Economic Behavior} \textbf{64}, 514-532.
\bibitem{25}Dickson A., Hartley R. (2013a). Bilateral oligopoly and quantity competition. \textit{Economic Theory} \textbf{52}, 979-1004.
\bibitem{26}Dickson A., Hartley R. (2013b). On ``nice'' and ``very nice" autarkic equilibria in strategic market games. \textit{Manchester School} \textbf{81}, 745–762.
\bibitem{27}Dr\`eze, J.H., Gabszewicz J.J., Gepts S. (1969). On cores and competitive equilibria. \textit{Aggregation et Dynamique des Ordres de Preference, Colloques Internationaux du C.N.R.S.} \textbf{171}, 91-114.
\bibitem{56}Dubey P., Shapley L.S. (1994). Noncooperative general exchange with a continuum of traders: two models. \textit{Journal of Mathematical Economics} \textbf{23}, 253-293.
\bibitem{28}Dubey P., Shubik, M. (1977). A closed economic system with production and exchange modelled a game of strategy. \textit{Journal of Mathematical Economics} \textbf{4}, 253-287.
\bibitem{29}Dubey P., Shubik, M. (1978). The noncooperative equilibria of a closed trading economy with market supply and bidding strategies. \textit{Journal of Economic Theory} \textbf{17}, 1-20.
\bibitem{30}Duffy J., Matros A., Temzelides T. (2011). Competitive behavior in market games: evidence and theory. \textit{Journal of Economic Theory} \textbf{146}, 1437–1463.
\bibitem{31} Edgeworth, F. Y. (1881). \textit{Mathematical Psychics}. Augustus M. Kelley.
\bibitem{32}Gabszewicz J.J. (2013). Introduction. \textit{Recherches \'Economiques de Louvain}, \textbf{79} 5-13.
\bibitem{33}Gabszewicz J.J., Mertens J.F. (1971). An equivalence theorem for
the core of an economy whose atoms are not ``too'' big. \textit{Econometrica} \textbf{39}, 713-721.
\bibitem{34}Gabszewicz J.J., Michel P. (1997). Oligopoly equilibrium in exchange economies. In Eaton B.C., Harris R.G. (eds), \textit{Trade, technology and economics: essays in honour of Richard G. Lipsey}, Edward Elgar, Cheltenham.
\bibitem{35}Gabszewicz J.J., Vial J.P. (1972). Oligopoly a l\`a Cournot in a general equilibrium analysis. \textit{Journal of Economic Theory} \textbf{4}, 381-400.
\bibitem{36}Gale D. (2000). Strategic Foundations of General Equilibrium: Dynamic Matching and Bargaining Games. Cambridge University Press.
\bibitem{37}Gary-Bobo R. (1988). Equilibre g\'en\'eral et concurrence imparfaite: un tour d'horizon. \textit{Recherches \'Economiques de Louvain} \textbf{54}, 53-84.
\bibitem{38} Groh, C. (1999). Sequential Moves and Comparative Statics in Strategic Market Games. \textit{Mimeo}. Department of Economics, University of Mannheim.
\bibitem{39}Hart O.D. (1985). Imperfect competition in general equilibrium: an overview of recent work. In Arrow K.J., Honkapohja S. (eds), \textit{Frontiers of Economics}, Basil Blackwell, Oxford.
\bibitem{40}Koutsougeras L.C., Meo C. (2017). An asymptotic analysis of strategic behavior for exchange economies. \textit{Economic Theory}
\bibitem{41}Lahmandi-Ayed R. (2001). Oligopoly equilibria in exchange economies:
a limit theorem. \textit{Economic Theory} \textbf{17}, 665–674.
\bibitem{42}Levando D. (2012). A survey of strategic market games. \textit{Economic Annals} \textbf{57}, 63-106.
\bibitem{43}Mas-Colell A. (1980). Noncooperative approaches to the theory of perfect competition: presentation. \textit{Journal of Economic Theory} \textbf{22}, 121-135.
\bibitem{54}Mas-Colell A. (1982). The Cournotian foundations of Walrasian equilibrium theory: an exposition of recent theory. In Hildenbrand W. (ed.), \textit{Advances in Economic Theory}, Cambridge University Press.
\bibitem{44} McKenzie L. W. (2002). \textit{Classical general equilibrium theory}. MIT press.
\bibitem{45}Okuno M., Postlewaite A., Roberts J. (1980). Oligopoly and competition in large markets. \textit{American Economic Review} \textbf{70}, 22-31.
\bibitem{46}Peck, J., Shell, K., Spear, S. (1992) The market game: existence and structure of equilibrium. J. Math. Econ. 21, 271–299.
\bibitem{47}Sahi S., Yao S. (1989). The noncooperative equilibria of a trading economy with complete markets and consistent prices. \textit{Journal of Mathematical Economics} \textbf{18}, 325-346.
\bibitem{48}Shapley L.S. (1976). Noncooperative general exchange. In Lin S.A.Y. (ed), \textit{Theory of Measurement of Economic Externalities}, Academic Press, New York.
\bibitem{49}Shapley L.S., Shubik M. (1977). Trade using one commodity as a means of payment. \textit{Journal of Political Economy} \textbf{85}, 937-968.
\bibitem{50}Shirai K. (2010). An existence theorem for Cournot–Walras equilibria in a monopolistically competitive economy. \textit{Journal of Mathematical Economics} \textbf{46}, 1093–1102.
\bibitem{51}Shitovitz B. (1973). Oligopoly in markets with a continuum of traders. \textit{Econometrica} \textbf{41}, 467-501.
\bibitem{52} Shubik M. (1973). Commodity, money, oligopoly, credit and bankruptcy in a general equilibrium model. \textit{Western Economic Journal} \textbf{11}, 24-38.
\bibitem{53}Varian H.R. (1992). \textit{Microeconomic Analysis} (3rd edition). W.W. Norton \& Company, New York.
\end{thebibliography}
\end{document}